\documentclass[10pt,conference]{IEEEtran}
\IEEEoverridecommandlockouts

\usepackage{graphicx}
\usepackage{amsmath}
\usepackage{amssymb}
\usepackage{booktabs}
\usepackage{enumitem}
\usepackage{tikz}
\usepackage{multirow}
\usepackage{caption}
\usetikzlibrary{arrows.meta,positioning,fit,calc}

\makeatletter
\newcommand{\linebreakand}{%
  \end{@IEEEauthorhalign}
  \hfill\mbox{}\par
  \mbox{}\hfill\begin{@IEEEauthorhalign}
}
\makeatother

\def\arxivdoi{}
\usepackage{eso-pic}
\AddToShipoutPictureBG*{%
  \AtPageLowerLeft{%
    \put(0,28){\makebox[\paperwidth]{\parbox{0.86\paperwidth}{\scriptsize\centering
      Accepted as a regular paper to the Industry \& Government Program of the 2026 IEEE International Conference on Big Data (IEEE BigData 2026).%
      \ifx\arxivdoi\empty\else\ DOI: \arxivdoi.\fi\\
      \copyright~2026 IEEE. Personal use of this material is permitted. Permission from IEEE must be
      obtained for all other uses, in any current or future media, including reprinting/republishing
      this material for advertising or promotional purposes, creating new collective works, for resale
      or redistribution to servers or lists, or reuse of any copyrighted component of this work in other works.}}}%
  }%
}

\begin{document}

\title{Unified Ranking Across Retrieval Channels: A Large-Scale E-Commerce Approach\thanks{$^{*}$These authors contributed equally to this work.}}

\author{
\IEEEauthorblockN{1\textsuperscript{st} Aditya Gaydhani$^{*}$}
\IEEEauthorblockA{\textit{Data Sciences} \\ \textit{Target Corporation}\\
Brooklyn Park, MN, USA \\
aditya.gaydhani@target.com}
\and
\IEEEauthorblockN{2\textsuperscript{nd} Guangyue Xu$^{*}$}
\IEEEauthorblockA{\textit{Data Sciences} \\ \textit{Target Corporation}\\
Brooklyn Park, MN, USA \\
guangyue.xu@target.com}
\and
\IEEEauthorblockN{3\textsuperscript{rd} Dhanush Kamath}
\IEEEauthorblockA{\textit{Data Sciences} \\ \textit{Target Corporation}\\
Sunnyvale, CA, USA \\
dhanush.kamath@target.com}
\linebreakand
\IEEEauthorblockN{4\textsuperscript{th} Ankit Singh}
\IEEEauthorblockA{\textit{Data Sciences} \\ \textit{Target Corporation}\\
Sunnyvale, CA, USA \\
ankit.singh2@target.com}
\and
\IEEEauthorblockN{5\textsuperscript{th} Alex Fengjie Li}
\IEEEauthorblockA{\textit{Data Sciences} \\ \textit{Target Corporation}\\
Sunnyvale, CA, USA \\
alex.li@target.com}
}

\maketitle

\begin{abstract}
Large-scale e-commerce search and recommendation systems rely on multiple specialized retrieval channels to surface diverse products, each optimized for a distinct objective. A key challenge is merging candidates from these heterogeneous channels into a single ranked list under strict latency constraints while optimizing business KPIs such as user conversion. Static rank-fusion methods such as Reciprocal Rank Fusion (RRF) and Weighted Interleaving use the same fusion behavior across queries, limiting their ability to adapt to context or model interactions among channel signals. We reformulate multi-channel fusion as a channel-aware learning-to-rank problem and propose a unified ranking model trained on a scalar, funnel-weighted graded-relevance label that combines clicks, add-to-carts, and purchases. The model incorporates channel-specific scores together with recent aggregate query--item behavior to capture near-term population demand. The evaluated channel set combines conventional relevance retrieval with query-conditioned, recommendation-oriented channels emphasizing newness, trending demand, regional relevance, and seasonality. A randomized online experiment shows a 2.85\% increase in user conversion over the deployed weighted-interleaving control. The model achieves a p95 reranking-stage latency below 50\,ms and has been deployed on Target.com.
\end{abstract}

\begin{IEEEkeywords}
big data, large-scale data processing, distributed systems, learning-to-rank, rank fusion, multi-channel retrieval, e-commerce search
\end{IEEEkeywords}

\section{Introduction}

Large-scale e-commerce systems consist of catalogs containing millions of products spanning diverse categories, price ranges, and life-cycle stages. Production search and recommendation systems operate as a multi-stage pipeline comprising retrieval, ranking, and re-ranking. Because it is computationally infeasible to score the entire catalog for every request under strict latency budgets, retrieval first selects a manageable candidate set of relevant products, after which downstream ranking stages refine the top-$k$ results to optimize user-facing metrics. To balance user satisfaction and business objectives, modern systems rely on multiple specialized retrieval channels, each optimized for a distinct objective (e.g., lexical relevance, semantic similarity, freshness, or seasonality)~\cite{comcastMCR2022, Huang_MultiChannelFusion_Retrieval}. In addition to lexical and semantic retrieval, channels emphasizing newness, trending demand, regional relevance, or seasonality act as query-conditioned recommendation-style candidate generators. They remain conditioned on the search query while emphasizing complementary discovery objectives. The final ranked list is produced by merging candidates from all of these heterogeneous channels.

Although multi-channel retrieval improves recall and diversity, it introduces significant re-ranking challenges. Channels are optimized independently and generate candidates with heterogeneous score distributions, bias characteristics, and objective functions. For instance, popularity-driven channels emphasize long-term engagement, whereas freshness-driven channels favor recency and exploration. Channel utility varies with query context and time, and the final ranking must optimize downstream business KPIs such as user conversion. These modeling challenges are further constrained by strict latency requirements, as even modest increases in response time can negatively impact user satisfaction and engagement~\cite{brutlag2009speed}. As a result, designing a fusion strategy that models context-dependent channel utility and cross-channel interactions under tight computational budgets remains technically challenging for e-commerce.

Prior static fusion methods such as RRF~\cite{rrf} and Weighted Interleaving~\cite{wi1,wi2} apply query-invariant fusion rules, limiting their ability to adapt channel influence to the query or exploit joint evidence across sources. Existing LTR approaches~\cite{Walmart2017,Kabir2024SurveyEcomLTR} are typically designed for single-channel candidate distributions and do not explicitly model heterogeneous retrieval sources. We address this gap with a unified, channel-aware LTR framework (Section~\ref{sec:problem_formulation}). We formulate multi-channel fusion for e-commerce search as a context-dependent learning-to-rank problem over heterogeneous query-conditioned candidate sources. Instead of relying on a single global fusion rule, the downstream ranker uses channel scores together with item and aggregate query--item features. The upstream channels retain their distinct objectives, while the final ranker optimizes one scalar business-aligned label that combines clicks, add-to-carts, and purchases. Recent aggregate behavioral signals reflect changing population-level demand rather than individual session intent. The resulting model places relevance-oriented and discovery-oriented channels under a common production reranker while meeting strict latency constraints in high-traffic production environments.

Our main contributions are summarized as follows:
\begin{itemize}
    \item We formulate and deploy a practical cross-channel ranking framework that replaces globally tuned fusion over independently ranked, query-conditioned channel outputs with a single learning-to-rank model.
    \item We propose a channel-aware representation over a fixed production channel set using each upstream channel's score and membership features.
    \item We introduce a weekly funnel-weighted graded-relevance label that scalarizes clicks, add-to-carts, and purchases into one conversion-aligned ranking objective.
    \item Through a large-scale online A/B experiment, we show a +2.85\% improvement in user conversion over the deployed weighted-interleaving control while meeting a p95 reranking-stage latency below 50\,ms.
\end{itemize}

\section{Related Work}
\label{sec:related_work}

\subsection{Rank Fusion and Multi-Channel Retrieval}
\label{sec:rw_fusion}
Reciprocal Rank Fusion (RRF)~\cite{rrf} and Weighted
Interleaving~\cite{wi1,wi2} are widely adopted strategies for merging
candidates from multiple retrieval channels. Standard RRF combines source ranks through a fixed reciprocal-rank transformation, whereas Weighted Interleaving samples from independently ranked lists using globally tuned channel-selection weights. Neither method conditions its fusion behavior on the query or models interactions among channel signals, so a channel that is highly informative for one query is weighted identically for every other query regardless of context. Huang et al.~\cite{Huang_MultiChannelFusion_Retrieval} systematically study
multi-channel fusion at the retrieval stage and show that optimized channel
weighting improves retrieval quality; however, their work focuses on
retrieval-level optimization and does not address how heterogeneous candidates
should be jointly modeled by a downstream ranking model. Our work addresses
this gap by learning a unified scoring function directly over the merged
candidate pool, treating channel membership and channel scores as learned features rather than as fixed combination weights.

\subsection{Learning to Rank in Industrial E-Commerce}
\label{sec:rw_ltr}
LTR models are central to industrial search
systems~\cite{Walmart2017,Kabir2024SurveyEcomLTR}. Despite rapid advances in
deep learning, tree-based models such as Gradient Boosted Decision Trees
(GBDT) remain widely deployed due to their efficiency and robustness under
realistic latency constraints~\cite{Huang_GBDT_vs_DL_LTR_RecSys2025,
googleGBDT_DNN}. Prior work~\cite{Amazon_2022_Clicks_As_Features} has demonstrated
the value of incorporating clicks as both training labels and input features;
we extend this idea to aggregate click, add-to-cart, and purchase signals in a multi-channel setting. Existing LTR work typically assumes a single homogeneous candidate distribution and does not explicitly model heterogeneous retrieval sources. In our setting, the upstream channels are optimized for distinct objectives, while the downstream LambdaMART model is trained on one scalar business-aligned objective that is shared across all channels at re-ranking time.

\subsection{System-Level Optimization across Ranking Stages}
\label{sec:rw_system}
Several works explore coordinated optimization across pipeline stages. Qin et
al.~\cite{Qin_MultiAgent_EcommerceSearch_SIGIR2022} propose a multi-agent
framework for e-commerce search that models interactions among retrieval and
ranking components. While these approaches emphasize cross-stage coordination,
they do not study unified ranking over heterogeneous multi-channel candidates
under strict production latency constraints. Our setting instead holds the upstream retrieval and channel-ranking logic fixed and focuses on the re-ranking stage, where the practical constraint is combining already-produced candidate lists under a p95 latency budget rather than re-optimizing the retrieval stage itself.

\subsection{Unified Modeling across Search and Recommendation}
\label{sec:rw_unified}
A related line of research investigates joint modeling of search and
recommendation. Liu et al.~\cite{Liu_Unified_Search_Recommendation_SIGIR2024}
propose a multi-scenario learning framework that shares representations across
search and recommendation by exploiting a global label space. Such methods
focus on representation learning and multi-task optimization, in line with broader work on learning transferable representations that generalize to unseen concepts~\cite{Xu_ZeroShotCCL_MetaNLP2021}, assuming a fixed
candidate generation process. Our work instead studies production search, where newness, trending, regional, and seasonal components provide query-conditioned recommendation-style candidates alongside lexical and semantic retrieval. This positions our channel set between conventional search retrieval and recommendation-style candidate generation, without requiring a shared model between the two surfaces.

\section{Methodology}

\subsection{Problem Formulation}
\label{sec:problem_formulation}

A large-scale e-commerce search system serves millions of queries daily across a catalog of tens of millions of products. Given a query $q \in \mathcal{Q}$,
the system retrieves and ranks items from a large corpus $\mathcal{I}$.
A single retrieval channel is insufficient to satisfy the diverse informational needs of users: lexical matching excels at exact queries, while semantic retrieval handles paraphrases and long-tail intent, including rare attribute--product combinations that require compositional generalization~\cite{Xu_GIPCOL_WACV2024}. The system therefore employs a set of $K$ heterogeneous retrieval channels
\begin{displaymath}
\mathcal{C} = \{c_1, c_2, \ldots, c_K\},
\end{displaymath}
where each channel $c_k$ is optimized for a distinct objective (e.g., lexical relevance, semantic similarity, item freshness, or long-term popularity).

\begin{figure*}[!t]
  \centering
  \includegraphics[width=0.9\textwidth]{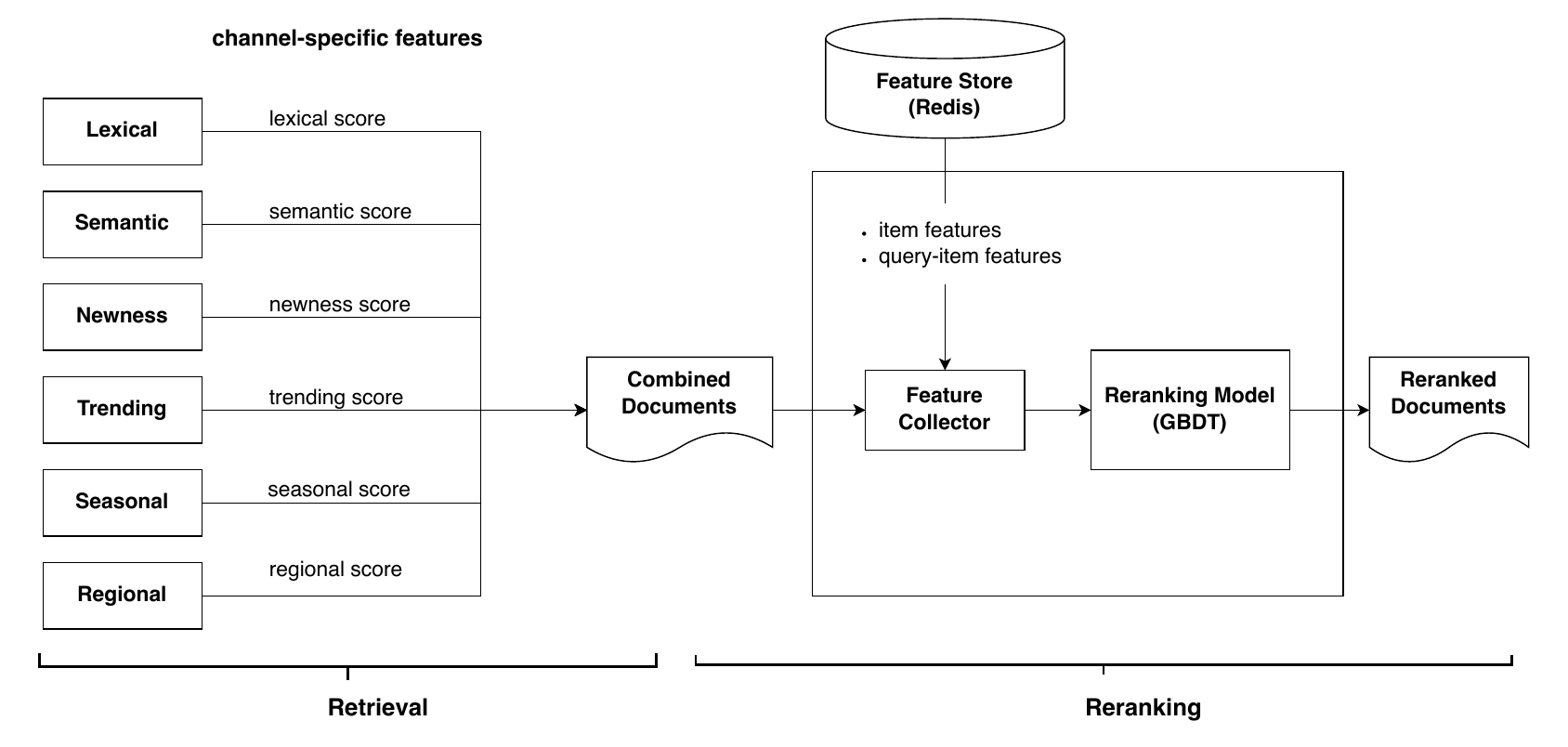}
  \caption{Overview of the unified ranking system. Multiple channel-specific retrieval components independently produce ranked candidate sets, which are merged and re-ranked by a single GBDT-based LTR model using channel-aware and engagement features.}
  \label{fig:arch}
\end{figure*}

For a given query $q$, each channel $c_k$ first applies its own production retrieval and ranking logic and independently produces a ranked list
\begin{displaymath}
\mathcal{R}_k(q) = \{(i, s_k(q, i)) \mid i \in \mathcal{I}\},
\end{displaymath}
where $s_k(q,i)$ denotes the channel-specific retrieval score. Because scoring all retrieved items in downstream stages is computationally infeasible, only the top-$n_k$ items from each channel are forwarded to re-ranking. We denote this truncated list as
\begin{displaymath}
\mathcal{R}_k^{(n_k)}(q) = \text{Top}_{n_k}\big(\mathcal{R}_k(q)\big).
\end{displaymath}

The final candidate pool for re-ranking is defined as the union of these truncated lists:
\begin{displaymath}
\mathcal{R}(q) = \bigcup_{k=1}^{K} \mathcal{R}_k^{(n_k)}(q).
\end{displaymath}
Note that items may appear in multiple channels; duplicates are deduplicated in $\mathcal{R}(q)$, with all channel-specific scores $\{s_k(q,i)\}$
retained as features for the re-ranking stage rather than being aggregated at this step.

The re-ranking stage learns a scoring function
\begin{displaymath}
f(q,i;\theta): \mathcal{Q} \times \mathcal{I} \rightarrow \mathbb{R},
\end{displaymath}
which assigns a unified relevance score to each $(q,i)$ with $i \in \mathcal{R}(q)$. Items are sorted in descending order of $f(q,i;\theta)$ to produce the final ranked list. The current model is trained and deployed for query-driven search and assumes a fixed production channel set $\mathcal{C}$ available during training. The scoring function $f$ is learned from historical interaction data; the following sections describe the training data representation, label construction, and features used in production.

\subsection{Data Representation}
\label{sec:data_representation}

All training instances are defined at the \emph{query--item--week} level to incorporate temporal dynamics into the ranking process. Let $t_w$ denote the beginning of week $w$. The label for $(q,i,w)$ is computed from interactions occurring during $[t_w,t_{w+1})$, while every behavioral feature for that instance is computed exclusively from events before $t_w$. For example, the 7-day feature window is $[t_w-7\text{ days},t_w)$, and the 14-day and 30-day windows end at the same cutoff. Feature and label windows therefore do not overlap.

Weekly aggregation was chosen empirically as a balance between responsiveness and statistical stability. Daily windows are too sparse and noisy for tail queries, while bi-weekly windows respond more slowly to seasonal, promotional, and emerging demand. Interaction counts and labels are reset at the beginning of each week rather than accumulated over the lifetime of a query--item pair. This limits the carryover of stale historical popularity and allows current population demand for new, trending, and seasonal products to affect the training target.

Modeling temporal structure is important in multi-channel retrieval settings, where channels may emphasize distinct objectives such as long-term popularity, short-term trendiness, or freshness. To capture both stable and emerging patterns, we incorporate behavioral features computed over multiple lookback windows ending at $t_w$~\cite{liu2024lookback}, including long-term aggregates (e.g., historical sales and engagement) and short-term signals that capture population-level demand velocity. This design limits reliance on lifetime popularity while allowing query-dependent, time-sensitive channel utility. It does not enforce channel-level exposure quotas. Given this temporal data structure, we next define the supervision signal used to train the ranking model.

\subsection{Label Construction}
\label{sec:label_construction}

User interactions in e-commerce search follow a conversion hierarchy:
\[
\text{Impression} \rightarrow \text{Click} \rightarrow \text{AddToCart} \rightarrow \text{Purchase}.
\]
For each query--item pair within a weekly window $(q,i,w)$, we summarize engagement over $N$ sessions using the deepest action per session.
Let $\{V_{q,i,w}, C_{q,i,w}, A_{q,i,w}, P_{q,i,w}\}$ denote counts of \emph{view-only}, \emph{click}, \emph{add-to-cart}, and \emph{purchase} outcomes, respectively, with $N = V_{q,i,w} + C_{q,i,w} + A_{q,i,w} + P_{q,i,w}$.
We define a scalar funnel-weighted graded-relevance label as a weighted aggregation:
\[
L(q,i,w) = a P_{q,i,w} + b A_{q,i,w} + c C_{q,i,w} + d V_{q,i,w},
\]
where $a \ge b \ge c \ge d \ge 0$. We calibrate weights using corpus-level
conversion statistics aggregated over the training period:
\[
a = 1, \quad
b = \frac{|\mathcal{P}|}{|\mathcal{A}|}, \quad
c = \frac{|\mathcal{P}|}{|\mathcal{C}|}, \quad
d = 0,
\]
where $|\mathcal{P}|$, $|\mathcal{A}|$, and $|\mathcal{C}|$ denote total purchases, add-to-carts, and clicks in the training corpus. Setting $d = 0$
excludes view-only impressions from the label, as they do not indicate purchase intent and would dilute the conversion signal---the primary business objective for Target's search system.

To ensure comparability across queries, we apply per-query max normalization:
\[
\tilde{L}(q,i,w) = \frac{4 \, L(q,i,w)}{\max_{i' \in \mathcal{R}(q,w)}
L(q,i',w)},
\]
mapping labels to $[0,4]$. The factor of 4 aligns with the standard graded relevance scale used in LambdaMART training, where higher values correspond
to more relevant items (e.g., $\tilde{L}=4$ for the most-purchased item per query).
We normalize at the query level rather than globally across all queries, as global normalization would conflate queries with vastly different engagement volumes and introduce high variance for sparse queries---a common challenge in long-tail e-commerce search traffic~\cite{exploitation_bias} and, more generally, in learning from limited examples~\cite{Xu_MetaReVision_EMNLP2023}.
Engagement counts and labels are reset for each weekly window rather than accumulated over the lifetime of the query--item pair. The target is therefore designed to optimize realized weekly business value rather than estimate a stationary per-impression conversion propensity. This limits long-horizon popularity carryover and allows newly released, trending, regional, or seasonal candidates to gain relevance when they attract current-week engagement and conversion, although it does not impose channel-specific exposure guarantees.
With labels defined, we describe the features used to represent each query--item--week instance.

\subsection{Features}
\label{sec:features}

We construct ranking features at a weekly granularity (Section~\ref{sec:data_representation}) and group them into three categories.

\paragraph{Item Features.}
These capture intrinsic product attributes (e.g., price, category, brand, availability) and behavioral aggregates over multiple lookback windows
(e.g., 7-day, 14-day, 30-day), all ending before the beginning of the label week. The features model long-term popularity, recency, and seasonality, enabling the model to balance historically strong performers with newly trending or seasonal products.

\paragraph{Channel-Aware Query--Item Features.}
For each query--item pair, channel-aware features include the raw retrieval score $s_k(q,i)$ from each channel. Because channel scores reflect distinct upstream objectives, these features provide the model with signals from which to learn query-dependent channel utility and cross-feature interactions. Items absent from a given channel receive \texttt{NA} values for the corresponding features, which are handled natively by the GBDT model without imputation.

\paragraph{Query--Item Engagement Features.}
Extending prior work on using clicks as both labels and features~\cite{Amazon_2022_Clicks_As_Features}, we incorporate clicks, add-to-cart events, and purchases as engagement features. These features are computed strictly before $t_w$ and use a weighted formulation similar to Section~\ref{sec:label_construction}, with distinct aggregation and temporal decay to capture short-term population demand velocity. For feature computation, data is aggregated across users and across all query variations identified by our query understanding models (e.g., spelling corrections and query rewrites), ensuring robust signal aggregation for semantically equivalent queries and improving coverage for head and torso queries. These signals capture aggregate query--item demand. Their impact is evaluated in Section~\ref{sec:rq2}.
These features are used to train a GBDT-based ranking model, described next.

\subsection{Model and Training}
\label{sec:model_and_training}

We adopt Gradient Boosted Decision Trees (GBDT) as the ranking model for three reasons specific to this production setting. First, GBDT handles
heterogeneous feature types---continuous retrieval scores, categorical channel indicators, and sparse engagement signals---without requiring feature
normalization. Second, it supports native \texttt{NA} handling for missing channel features, eliminating the need for imputation. Third, it provides deterministic and efficient inference, critical for meeting the p95 $<$50\,ms reranking-stage latency requirement under high-traffic production conditions~\cite{Huang_GBDT_vs_DL_LTR_RecSys2025,GBDT2001}.
In large-scale production search systems, tree-based models remain competitive with deep neural approaches under realistic latency and feature constraints, particularly when feature engineering plays a central role~\cite{Huang_GBDT_vs_DL_LTR_RecSys2025, googleGBDT_DNN}. GBDT constructs an additive ensemble of decision trees:
\begin{displaymath}
f(q, i; \theta) = \sum_{t=1}^{T} \eta \cdot h_t(q, i),
\end{displaymath}
where $h_t$ is the $t$-th tree, $T$ is the total number of trees, and $\eta$ is the learning rate (shrinkage factor). Each tree is trained sequentially to minimize the residual loss from previous iterations. Our model is implemented using the Yggdrasil Decision Forests library~\cite{YDF}, which provides efficient training and inference optimized for high-throughput serving environments.

We train the model using the LambdaMART objective~\cite{lambdamart}, whose pairwise gradient updates are weighted by changes in NDCG. This aligns naturally with our graded relevance labels $\tilde{L} \in [0,4]$ derived from the conversion hierarchy and enables the model to prioritize high-value purchase events while still accounting for clicks and add-to-cart signals.

To effectively model heterogeneous retrieval signals and behavioral features, we employ a local tree growing strategy with sparse oblique splits (linear
combinations of features at each split node), enabling efficient handling of correlated and high-dimensional feature spaces such as multi-channel
retrieval scores. Regularization is applied through L2 penalties and shrinkage to control model complexity and mitigate overfitting. We enable second-order (Hessian-based) gain computation to improve split quality and stabilize training.
Hyperparameters---including tree depth, number of trees, learning rate (shrinkage), and minimum leaf size---are tuned via grid search on the
held-out fourth-week validation set described in Section~\ref{sec:dataset}. Final model selection is based on offline NDCG@8 on the validation set.

\section{Experiment Design}

All experiments in this paper are conducted in an e-commerce search system. The evaluated candidate pool combines lexical and semantic retrieval with channels that emphasize newness, trending demand, regional relevance, and seasonality while remaining conditioned on the search query. These channels play a recommendation-like discovery role within the search pipeline. The results therefore concern a common reranker over heterogeneous query-conditioned channels, not a model shared between search and recommendation surfaces.
\subsection{Dataset}
\label{sec:dataset}

Our experiments are conducted using large-scale user interaction logs collected from Target's e-commerce search system, in accordance with internal privacy, security, and data governance policies. The dataset includes search queries, ranked item impressions, and corresponding user interactions, including clicks, add-to-cart events, and purchases, as defined in Section~\ref{sec:label_construction}. We use a recent five-week period to construct labeled examples. The dataset is partitioned chronologically: the first three label weeks are used for training, the fourth label week for validation and hyperparameter tuning, and the final label week is held out for offline evaluation. Behavioral feature windows may draw on historical events preceding this five-week label period, but always end strictly before the corresponding label week. This temporal split enables evaluation under concept drift and non-stationarity without overlapping feature and label windows.

To improve data quality and reduce noise from extreme long-tail queries, we filter the data at the query-week level, retaining only queries for which at least one item receives at least one purchase within the same week. For each retained query, we include up to the top-$n_k$ items per channel for reranking, as described in Section~\ref{sec:problem_formulation}, to prevent train--inference skew. The final training set comprises approximately 60M rows and 500k unique queries distributed across head, torso, and tail segments. Queries are segmented using production query-frequency thresholds into head, torso, and tail groups, each accounting for 33\% of search traffic.

From a data-processing perspective, this representation creates a high-cardinality aggregation problem: the same event stream must be grouped along query, item, and temporal dimensions while preserving multiple historical lookback windows and enforcing point-in-time correctness. The approximately 60M-row data is therefore generated as a distributed dataset rather than as an in-memory modeling table. This organization is important operationally because feature computation grows with interaction volume and the number of query--item keys, rather than only with the number of unique catalog items.

Labels and features are computed for each query--item--week instance as described in Sections~\ref{sec:label_construction} and~\ref{sec:features}, respectively.

\subsection{Model Variants}
\label{sec:variants}

We evaluate four model variants to isolate the contribution of each component. UR and UR + EF use the same production graded-relevance label, whose coefficients are determined heuristically based on internal business requirements. Holding that label fixed allows the comparison to isolate the addition of engagement features.

\begin{itemize}[leftmargin=*]
  \item \textbf{WI (Baseline).} Each channel first applies its own production retrieval and ranking logic using channel-specific features and objectives. The deployed Weighted Interleaving layer then constructs the final ranking by probabilistically sampling from these independently ranked lists according to globally tuned channel-selection weights~\cite{wi1,wi2}. We use this system as the control because it is the production fusion mechanism being replaced; an exhaustive benchmark of static rank-fusion algorithms is outside the scope of this work.

  \item \textbf{UR.} Unified ranking using the same upstream channel outputs and a GBDT model trained on item features and channel-aware query--item features (Section~\ref{sec:features}). This variant replaces only the final weighted-interleaving fusion layer with a learned cross-channel reranker.

  \item \textbf{UR + EF.} Unified ranking with engagement features. This variant extends UR by incorporating aggregate query--item clicks, add-to-cart events, and purchases over multiple pre-week time windows (Section~\ref{sec:features}). It uses the same production label as UR, so the comparison isolates the contribution of these behavioral features.

  \item \textbf{UR + EF + CL.} Unified ranking with engagement features and the conversion-weighted label. This is the full proposed model. It retains the architecture and features of UR + EF but replaces the production label with the scalar funnel-weighted label described in Section~\ref{sec:label_construction}.
\end{itemize}

\subsection{A/B Test Setup}
\label{sec:ab_test_setup}

We evaluated all ranking variants through a multivariate online A/B experiment on the production search platform. Traffic was randomly assigned at the visitor level, with each variant receiving 5\% of eligible search traffic. The experiment ran for two full weeks and included all users, with bot traffic filtered out. The evaluated variants were WI (control), UR, UR + EF, and UR + EF + CL. We report relative lifts over the control for click-through rate (CTR), add-to-cart rate (ATC), and conversion rate. Statistical significance was assessed using a two-sided Student's $t$-test at the 95\% confidence level. All significance markers compare an individual treatment arm with the WI control. Prior to the experiment, a power analysis assuming 80\% power, a 95\% confidence level, and a 1\% minimum detectable effect in conversion estimated the required sample size for significance.

\section{Results}
\label{sec:results}

Table~\ref{tab:overall_results} summarizes the offline and online performance of all model variants. Offline NDCG@8 is used as a development and model-selection metric against held-out behavior-derived labels as the top-$8$ slots receive majority of the user engagement. Conclusions about production impact are based on the randomized online A/B experiment. We discuss each ablation in turn, alongside qualitative examples illustrating the corresponding changes in search ranking.

\begin{table}[!ht]
  \caption{Offline and online performance comparison of model variants. Online metrics are reported as lift (\%) over the weighted-interleaving control. * indicates statistical significance at the 95\% confidence level for the corresponding treatment-versus-WI comparison.}
  \label{tab:overall_results}
  \centering
  \renewcommand{\arraystretch}{1.4}
  \setlength{\tabcolsep}{6pt}
  \begin{tabular}{l c | c c c}
    \toprule
     & \multicolumn{1}{c|}{\textbf{Offline}} & \multicolumn{3}{c}{\textbf{Online (A/B Test Lift \%)}} \\
    \textbf{Model Variant}
    & \textbf{NDCG@8}
    & \textbf{CTR}
    & \textbf{ATC}
    & \textbf{Conversion} \\
    \midrule
    WI (Baseline) & 0.6620  & -- & -- & -- \\
    UR & 0.7169 & +0.26 & +1.21* & +1.28* \\
    UR + EF & 0.7799 & +1.52* & +2.72* & +2.38* \\
    UR + EF + CL & 0.7994 & +1.46* & +2.81* & +2.85* \\
    \bottomrule
  \end{tabular}
\end{table}

\subsection{Weighted Interleaving vs.\ Unified Ranking}
\label{sec:rq1}

We evaluate the unified ranking model against the weighted interleaving baseline shown in Table~\ref{tab:overall_results}. In the baseline, independently ranked channel outputs are probabilistically merged according to globally tuned channel weights~\cite{wi1, wi2}. Unified ranking receives the same upstream channel outputs but replaces only the final fusion layer with a query--item scoring function (Section~\ref{sec:problem_formulation}) aligned with the downstream objective.

As reported in Table~\ref{tab:overall_results}, unified ranking increases offline NDCG@8 from 0.6620 to 0.7169, indicating closer agreement with the behavior-derived development objective. Relative to WI, the online A/B test shows positive point estimates across all business metrics and statistically significant improvements in add-to-cart (+1.21\%) and conversion (+1.28\%). These findings show that replacing the deployed globally tuned fusion layer with a learned cross-channel reranker yields measurable production impact.

\begin{figure*}[!t]
  \centering
  \includegraphics[width=0.9\textwidth]{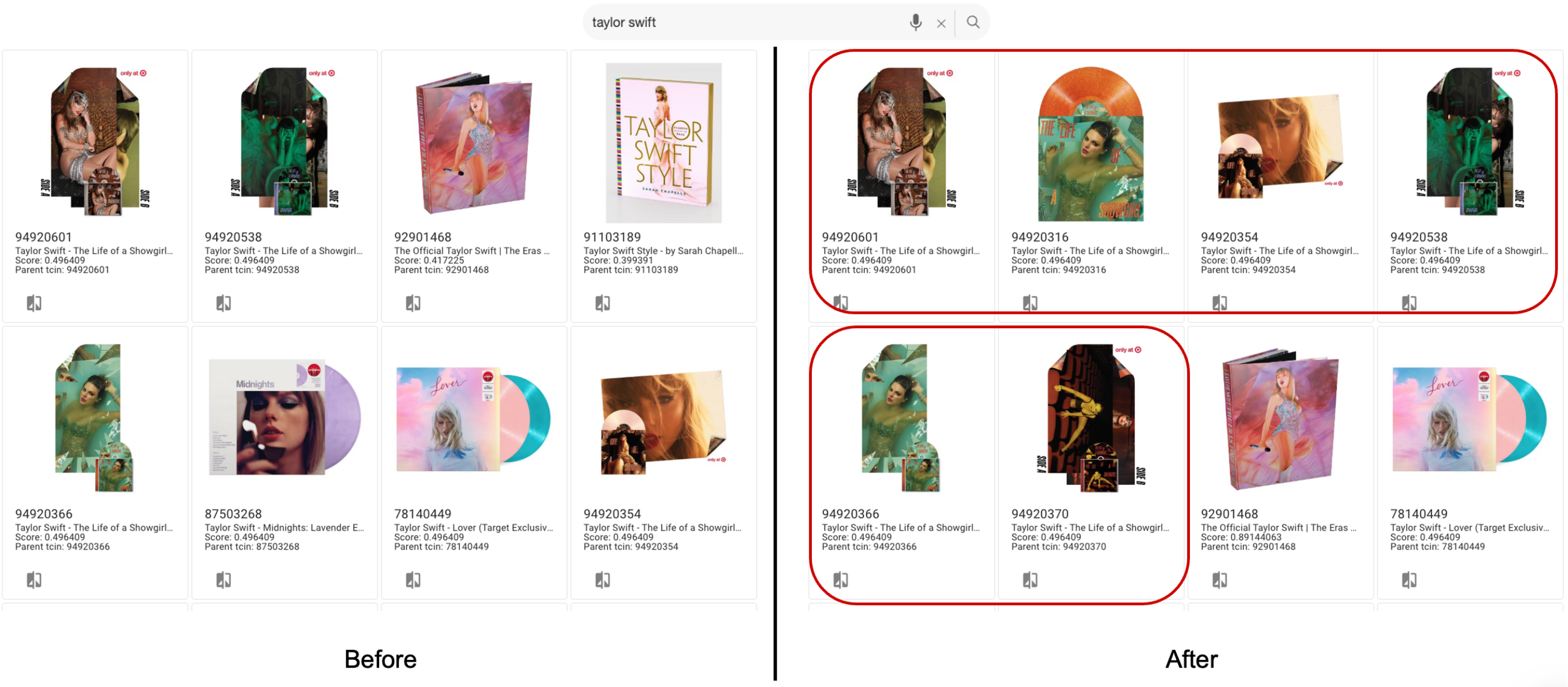}
  \caption{Top 8 ranked results for the query \emph{taylor swift}. The results on the right surface the latest releases by Taylor Swift at the top of the grid because newer items show higher engagement in this category.}
  \label{fig:example_taylor_swift}
\end{figure*}

\begin{figure*}[!t]
  \centering
  \includegraphics[width=0.9\textwidth]{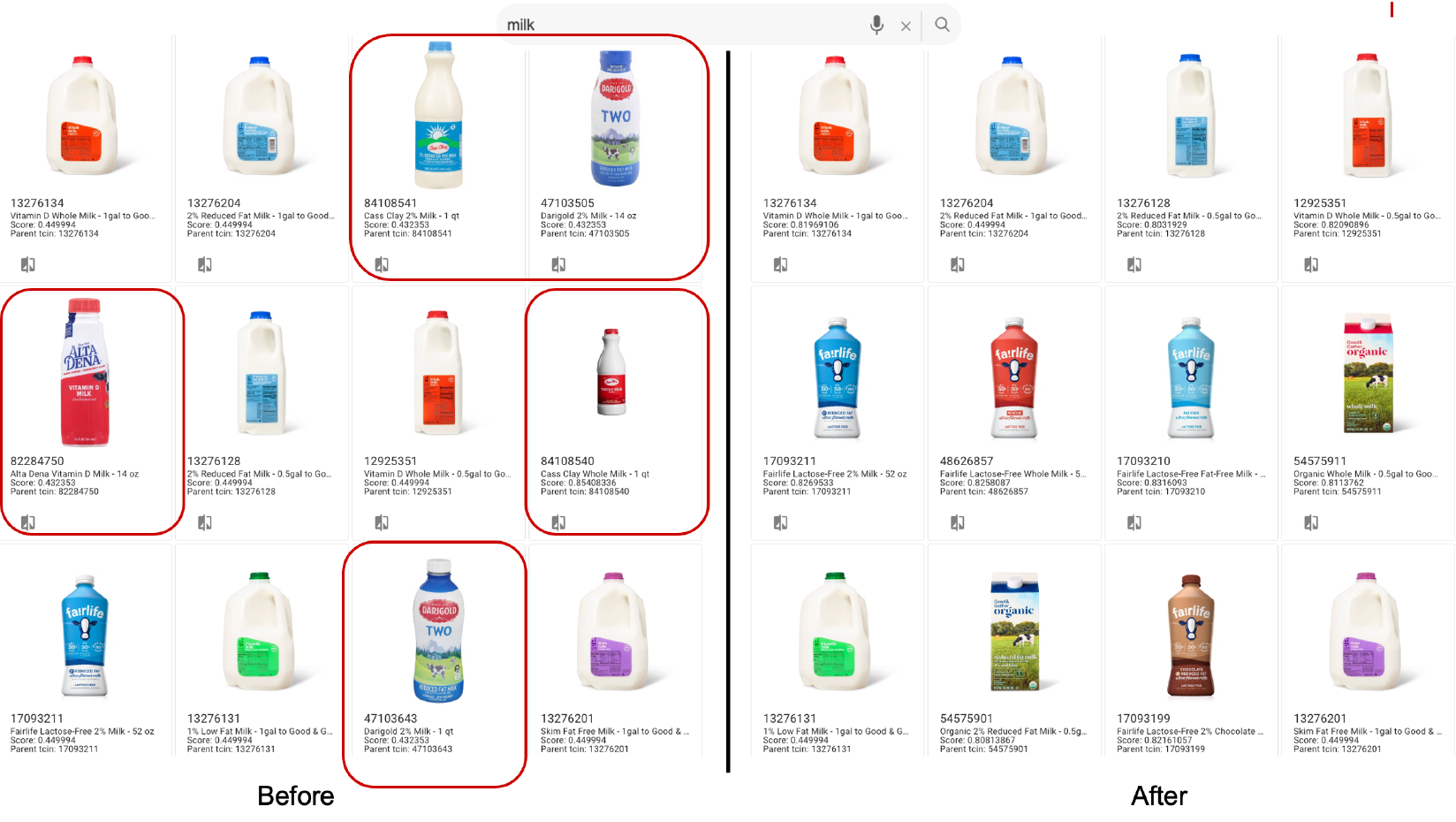}
  \caption{Top 12 ranked results for the query ``milk''. The highlighted items on the left are produced by weighted interleaving and exibit sub-par performance compared to the results on the right, which are driven by the unified model.}
  \label{fig:example_milk}
\end{figure*}

Figure~\ref{fig:example_taylor_swift} and Figure~\ref{fig:example_milk} highlight the qualitative difference between weighted interleaving on the left and unified ranking on the right. The unified model places recent high-engagement products nearer the top of the ranking.

\subsection{Contribution of User Engagement Features}
\label{sec:rq2}

We evaluate the contribution of query--item engagement features (Section~\ref{sec:features}) through an ablation study comparing the unified ranking model with and without these features.

As shown in Table~\ref{tab:overall_results}, UR + EF increases offline NDCG@8 from 0.7169 to 0.7799. Relative to WI, it produces statistically significant lifts in CTR (+1.52\%), add-to-cart (+2.72\%), and conversion (+2.38\%). These online point estimates are larger than those observed for UR and support the value of aggregate engagement features. The result indicates that historical query--item interactions provide a strong predictive signal for current population-level demand.

\begin{figure*}[!t]
  \centering
  \includegraphics[width=0.9\textwidth]{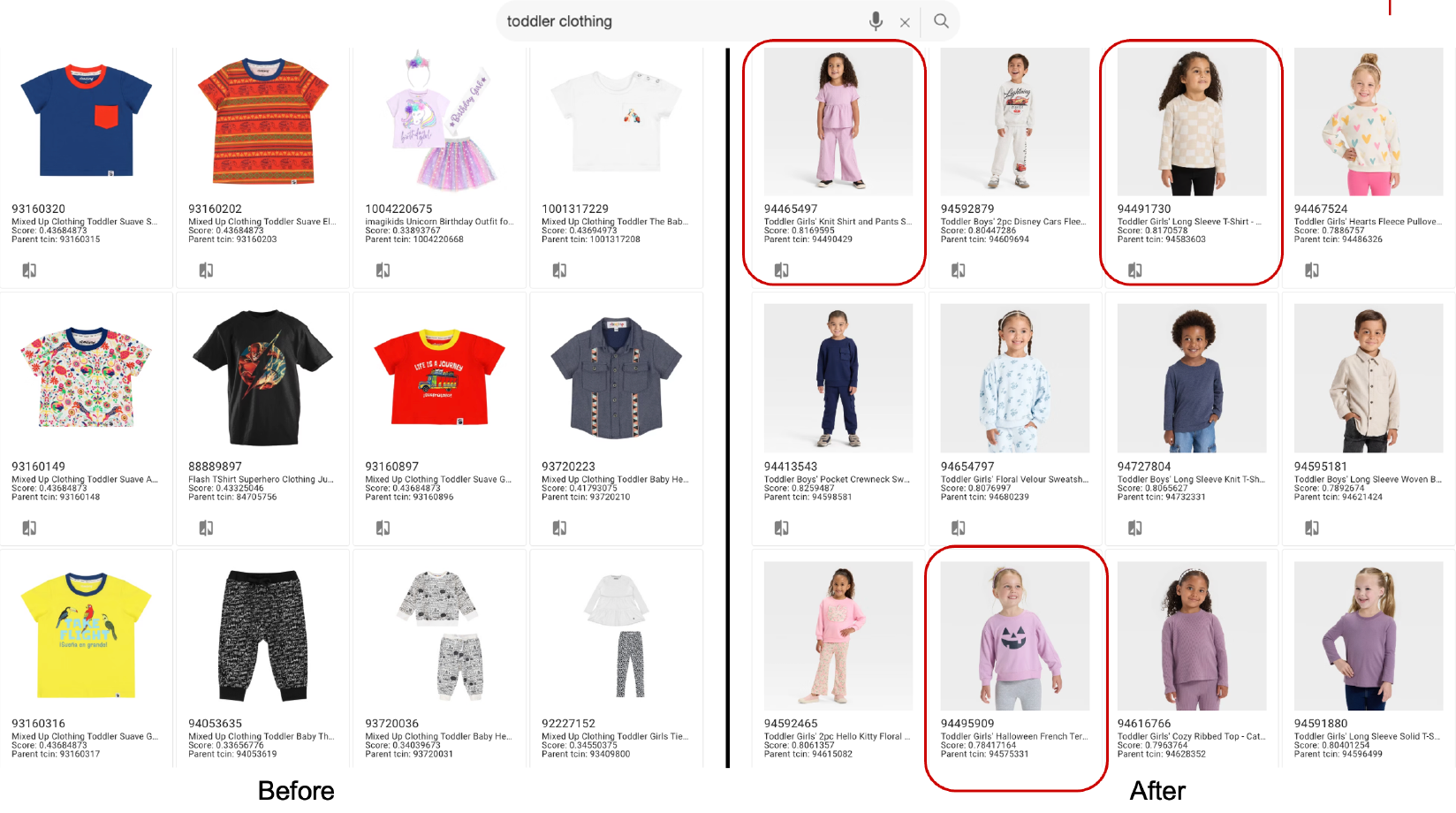}
  \caption{Top 12 ranked results for the query \emph{toddler clothing}. The results after the model update show improved engagement-aware ordering and broader semantic retrieval coverage.}
  \label{fig:example_toddler_clothing}
\end{figure*}

Figure~\ref{fig:example_toddler_clothing} highlights the contribution of aggregate engagement features. Popular products from the semantic channel appear nearer the top of the grid, whereas the model on the left lacks these signals and relies more heavily on the upstream ordering. The unified model also surfaces a seasonal Halloween-themed product during October that is absent from the comparison ranking's top 12.

\subsection{Effectiveness of Conversion-Weighted Engagement Label}
\label{sec:rq3}

To isolate the impact of label construction, we compare a unified ranking model trained with the heuristic production label against an otherwise identical model trained with the proposed scalar conversion-weighted label (Section~\ref{sec:label_construction}).

Table~\ref{tab:overall_results} shows that UR + EF + CL attains an offline NDCG@8 of 0.7994, compared with 0.7799 for UR + EF. Relative to WI, the full model produces statistically significant lifts in add-to-cart (+2.81\%) and conversion (+2.85\%) while maintaining a similar CTR lift. It has the highest observed conversion point estimate among the evaluated treatments. The findings indicate that scalar weighting of engagement outcomes across the conversion funnel aligns model training with downstream business value.

\subsection{Query Segment Analysis}
\label{sec:query_segment_analysis}

\begin{table}[!ht]
\caption{Segment-level A/B test performance. Metrics are reported as lift (\%) over the weighted-interleaving control. * indicates statistical significance at the 95\% confidence level for the corresponding treatment-versus-WI comparison.}
\label{tab:segment_results}
\centering
\renewcommand{\arraystretch}{1.2}
\setlength{\tabcolsep}{6pt}
\begin{tabular}{l l c c c}
\toprule
\textbf{Model} & \textbf{Query Segment} & \textbf{CTR} & \textbf{ATC} & \textbf{Conversion} \\
\midrule
\multirow{3}{*}{UR}
& Head  & +0.76 & +1.95* & +2.23* \\
& Torso & -1.09 & +1.35* & +1.22* \\
& Tail  & -1.13 & +1.11* & +1.91* \\
\midrule
\multirow{3}{*}{UR + EF}
& Head  & +2.70* & +4.28* & +4.39* \\
& Torso & +0.52 & +2.27* & +2.87* \\
& Tail  & -0.21 & +1.93* & +1.89* \\
\midrule
\multirow{3}{*}{UR + EF + CL}
& Head  & +2.92* & +4.76* & +5.27* \\
& Torso & -2.25* & +1.06* & +2.56* \\
& Tail  & -2.23* & +0.38 & +1.88* \\
\bottomrule
\end{tabular}
\end{table}

Table~\ref{tab:segment_results} reports online performance across head, torso, and tail query segments. All three unified-ranking variants produce statistically significant conversion lifts in every segment. Add-to-cart lifts are also statistically significant in every case except UR + EF + CL on tail queries (+0.38\%). For head queries, UR + EF and UR + EF + CL significantly improve CTR, add-to-cart, and conversion, and the full model has the largest observed conversion lift (+5.27\%). For torso and tail queries, UR + EF + CL produces significant CTR declines (-2.25\% and -2.23\%, respectively) while retaining significant conversion lifts (+2.56\% and +1.88\%). This pattern is consistent with the scalar conversion-weighted objective prioritizing downstream funnel outcomes even when CTR decreases.

\section{Deployment and Serving}
\label{sec:deployment}

The unified ranking model is deployed within the production search stack, which combines distributed offline data processing for training with a low-latency online serving path.

\subsection{Training Pipeline}
\label{sec:deployment_training}
We train Gradient Boosted Decision Tree (GBDT) ranking models \cite{GBDT2001} using Yggdrasil Decision Forests (YDF) \cite{YDF}. Offline feature pipelines built on Spark \cite{spark} generate the training data at the query--item--week granularity described in Section~\ref{sec:data_representation}, and a batch training job trains the YDF model before publishing the resulting artifact to Google Cloud Storage (GCS) \cite{gcp} for downstream serving.

The offline computation is organized as a sequence of distributed transformations over interaction logs rather than a monolithic model-training job. Event records are first normalized to production query identifiers and joined with item and channel metadata. The pipeline then performs keyed temporal aggregations to construct click, add-to-cart, purchase, and item-level statistics for each required lookback window. Labels are computed from the subsequent weekly window only after the historical feature cutoffs have been materialized. This separation makes point-in-time correctness explicit and avoids leakage when processing large event histories in parallel. Intermediate datasets are partitioned by temporal and query-derived keys so that weekly examples can be recomputed without rebuilding the entire history.

At this scale, the dominant engineering challenge is not the GBDT fit itself but the repeated aggregation and joining of sparse behavioral signals across high-cardinality query--item pairs. Spark provides data-parallel execution for these operations and allows the pipeline to scale horizontally as interaction volume grows. The resulting data is approximately 60M rows; after distributed feature construction, model training consumes the compact representation rather than raw event logs. This division of labor keeps large-scale data preparation in the distributed processing layer while retaining an inference-efficient model for serving.

\subsection{Scalability and Data Engineering}
\label{sec:scalability}
The architecture separates scale along three dimensions: corpus scale, interaction-data scale, and request-time candidate scale. Corpus scale is handled upstream by specialized retrieval systems, so the unified ranker never scans the tens-of-millions-item catalog online. Instead, each channel emits a bounded top-$n_k$ list and the ranker operates on the deduplicated union $\mathcal{R}(q)$. Consequently, reranking work is bounded by $|\mathcal{R}(q)| \leq \sum_k n_k$ before deduplication, decoupling online scoring cost from total catalog size. Adding a retrieval channel increases the candidate and feature dimensions in a controlled manner rather than requiring a full-corpus scoring pass.

Interaction-data scale is handled offline through distributed aggregation and multi-resolution temporal windows. Rather than storing raw histories in the online path, the pipeline compresses event streams into query--item and item-level statistics that are refreshed at real-time, daily, or weekly cadences according to how quickly each signal changes. This is a practical big-data tradeoff: expensive scans, joins, and windowed aggregations are performed asynchronously, while request-time computation is reduced to bounded key-value retrieval, feature assembly, and tree inference. The same feature definitions and temporal cutoffs are used across offline generation and online materialization to reduce training--serving skew.

The serving design also isolates components with different scaling characteristics. Candidate retrieval can fan out across channels, feature ingestion can scale independently through Kafka consumers, Redis provides horizontally scalable low-latency access to materialized features, and TensorFlow Serving hosts the model behind a stateless scoring interface. This decomposition permits independent capacity planning for retrieval throughput, feature-update throughput, feature-store read load, and model inference. It also provides graceful operational boundaries: slower-changing aggregates can be refreshed in batch without blocking real-time signals, and model artifacts can be published independently of feature recomputation.

\subsection{Online Serving Infrastructure}
\label{sec:deployment_serving}
For online serving, the production model is loaded and served through TensorFlow Serving \cite{tfserve}. Features are organized by update frequency into real-time, daily, and weekly feature sets. These features are materialized in Redis \cite{redis} to support low-latency lookup at request time. Real-time updates are consumed from Kafka topics \cite{kafka} and continuously written to Redis, while daily and weekly features are refreshed by scheduled batch jobs that update the same serving store, keeping the online feature store consistent with the offline training data despite the differing refresh cadences.

At request time, the Search Orchestration layer retrieves candidates from the available retrieval channels and passes them to the Ranking Orchestration layer. The Ranking Orchestration layer retrieves the required features from Redis, performs feature processing, calls the model-serving endpoint to score each candidate, applies business rules, and returns the final ranked response. In production traffic, the model achieves a p95 end-to-end latency of less than 50\,ms, measured from feature retrieval through scoring computation across all candidates in $\mathcal{R}(q)$.

\section{Conclusion and Future Work}

We introduced a channel-aware learning-to-rank framework that replaces globally tuned fusion over a fixed set of independently ranked retrieval channels with a single downstream reranker. The model combines channel scores with aggregate pre-week behavioral features and optimizes a scalar weekly funnel-weighted label. Relative to the deployed weighted-interleaving control, the randomized online experiment produced statistically significant lifts in add-to-cart and conversion for the full model, including a 2.85\% conversion lift, while maintaining a p95 reranking-stage latency below 50\,ms.

We view the candidate-union and reranking abstraction as a search-side building block toward future unified search and recommendation. Future work includes adapting the framework to recommendation surfaces with surface-specific retraining, systematically evaluating bias and exposure across channels, and incorporating individual personalization signals beyond the population-level demand features studied here, as well as richer semantic item representations that generalize to novel attribute--product compositions~\cite{Xu_PromptingVLM_CCL_arXiv2022,Xu_GroundedCCL_PhD2024}.

\section*{Acknowledgment}
We thank Search --- Data Science leadership at Target for their support in enabling this initiative. We are grateful to the Search Product team for their collaboration in formulating the case for incorporating user engagement features into the ranking framework, contributing to the A/B test design, and conducting the post-experiment analysis. We also thank the Query Understanding team for their work on query normalization, which improved the quality of the engagement features used in this study. Finally, we acknowledge the foundational contributions of the Search Retrieval and Engineering groups in developing the retrieval channels and ranker service that made this work possible, as well as the support of the Content Understanding and Search Platform teams.

\bibliographystyle{IEEEtran}
\bibliography{main}

\begin{thebibliography}{10}
\providecommand{\url}[1]{#1}
\csname url@samestyle\endcsname
\providecommand{\newblock}{\relax}
\providecommand{\bibinfo}[2]{#2}
\providecommand{\BIBentrySTDinterwordspacing}{\spaceskip=0pt\relax}
\providecommand{\BIBentryALTinterwordstretchfactor}{4}
\providecommand{\BIBentryALTinterwordspacing}{\spaceskip=\fontdimen2\font plus
\BIBentryALTinterwordstretchfactor\fontdimen3\font minus
  \fontdimen4\font\relax}
\providecommand{\BIBforeignlanguage}[2]{{%
\expandafter\ifx\csname l@#1\endcsname\relax
\typeout{** WARNING: IEEEtran.bst: No hyphenation pattern has been}%
\typeout{** loaded for the language `#1'. Using the pattern for}%
\typeout{** the default language instead.}%
\else
\language=\csname l@#1\endcsname
\fi
#2}}
\providecommand{\BIBdecl}{\relax}
\BIBdecl

\bibitem{comcastMCR2022}
\BIBentryALTinterwordspacing
S.~Rome, S.~Hamidian, R.~Walsh, K.~Foley, and F.~Ture, ``Learning to rank
  instant search results with multiple indices: A case study in search
  aggregation for entertainment,'' in \emph{Proceedings of the 45th
  International ACM SIGIR Conference on Research and Development in Information
  Retrieval}, ser. SIGIR '22.\hskip 1em plus 0.5em minus 0.4em\relax New York,
  NY, USA: Association for Computing Machinery, 2022, p. 3412–3416. [Online].
  Available: \url{https://doi.org/10.1145/3477495.3536334}
\BIBentrySTDinterwordspacing

\bibitem{Huang_MultiChannelFusion_Retrieval}
J.~Huang, J.~Qin, J.~Lin, Z.~Feng, Y.~Yu, and W.~Zhang, ``Unleashing the
  potential of multi-channel fusion in retrieval for personalized
  recommendations,'' \emph{arXiv preprint}, 2025.

\bibitem{brutlag2009speed}
J.~Brutlag, ``Speed matters for google web search,'' Google Inc., Tech. Rep.,
  Jun. 2009, google Technical Report.

\bibitem{rrf}
\BIBentryALTinterwordspacing
G.~V. Cormack, C.~L.~A. Clarke, and S.~Buettcher, ``Reciprocal rank fusion
  outperforms condorcet and individual rank learning methods,'' in
  \emph{Proceedings of the 32nd International ACM SIGIR Conference on Research
  and Development in Information Retrieval}, ser. SIGIR '09.\hskip 1em plus
  0.5em minus 0.4em\relax New York, NY, USA: Association for Computing
  Machinery, 2009, p. 758–759. [Online]. Available:
  \url{https://doi.org/10.1145/1571941.1572114}
\BIBentrySTDinterwordspacing

\bibitem{wi1}
Y.~Yue, Y.~Gao, O.~Chapelle, Y.~Zhang, and T.~Joachims, ``Learning more
  powerful test statistics for click-based retrieval evaluation,'' in
  \emph{Proceedings of the 33rd International ACM SIGIR Conference on Research
  and Development in Information Retrieval (SIGIR '10)}.\hskip 1em plus 0.5em
  minus 0.4em\relax ACM, 2010, pp. 507--514.

\bibitem{wi2}
K.~Hofmann, S.~Whiteson, and M.~de~Rijke, ``A probabilistic method for
  inferring preferences from clicks,'' in \emph{Proceedings of the 20th ACM
  International Conference on Information and Knowledge Management (CIKM
  '11)}.\hskip 1em plus 0.5em minus 0.4em\relax ACM, 2011, pp. 249--258.

\bibitem{Walmart2017}
\BIBentryALTinterwordspacing
S.~K. Karmaker~Santu, P.~Sondhi, and C.~Zhai, ``On application of learning to
  rank for e-commerce search,'' in \emph{Proceedings of the 40th International
  ACM SIGIR Conference on Research and Development in Information Retrieval},
  ser. SIGIR '17.\hskip 1em plus 0.5em minus 0.4em\relax New York, NY, USA:
  Association for Computing Machinery, 2017, p. 475–484. [Online]. Available:
  \url{https://doi.org/10.1145/3077136.3080838}
\BIBentrySTDinterwordspacing

\bibitem{Kabir2024SurveyEcomLTR}
\BIBentryALTinterwordspacing
M.~A. Kabir, M.~A. Hasan, A.~Mandal, D.~Tunkelang, and Z.~Wu, ``A survey on
  e-commerce learning to rank,'' \emph{arXiv preprint arXiv:2412.03581}, 2024.
  [Online]. Available: \url{https://arxiv.org/abs/2412.03581}
\BIBentrySTDinterwordspacing

\bibitem{Huang_GBDT_vs_DL_LTR_RecSys2025}
J.~Huang, J.~Qin, W.~Zhang, and Y.~Yu, ``Industry insights from comparing deep
  learning and gbdt models for e-commerce learning-to-rank,'' in
  \emph{Proceedings of the 19th ACM Conference on Recommender Systems
  (RecSys)}, 2025, industry Track.

\bibitem{googleGBDT_DNN}
Z.~Qin, L.~Yan, H.~Zhuang, Y.~Tay, R.~K. Pasumarthi, X.~Wang, M.~Bendersky, and
  M.~Najork, ``Are neural rankers still outperformed by gradient boosted
  decision trees?'' in \emph{International Conference on Learning
  Representations (ICLR)}, 2021.

\bibitem{Amazon_2022_Clicks_As_Features}
\BIBentryALTinterwordspacing
T.~Yang, C.~Luo, H.~Lu, P.~Gupta, B.~Yin, and Q.~Ai, ``Can clicks be both
  labels and features? unbiased behavior feature collection and
  uncertainty-aware learning to rank,'' in \emph{Proceedings of the 45th
  International ACM SIGIR Conference on Research and Development in Information
  Retrieval}, ser. SIGIR '22.\hskip 1em plus 0.5em minus 0.4em\relax New York,
  NY, USA: Association for Computing Machinery, 2022, p. 6–17. [Online].
  Available: \url{https://doi.org/10.1145/3477495.3531948}
\BIBentrySTDinterwordspacing

\bibitem{Qin_MultiAgent_EcommerceSearch_SIGIR2022}
J.~Qin, J.~Zhu, B.~Chen, Z.~Liu, W.~Liu, R.~Tang, R.~Zhang, Y.~Yu, and
  W.~Zhang, ``Beyond retrieval-ranking: A multi-agent cognitive decision
  framework for e-commerce search,'' in \emph{Proceedings of the 45th
  International ACM SIGIR Conference}, 2022.

\bibitem{Liu_Unified_Search_Recommendation_SIGIR2024}
J.~Liu, Q.~Chen, J.~Xu, J.~Li, B.~Li, and S.~Xu, ``A unified search and
  recommendation framework based on multi-scenario learning for ranking in
  e-commerce,'' in \emph{Proceedings of the 47th International ACM SIGIR
  Conference}, 2024.

\bibitem{Xu_ZeroShotCCL_MetaNLP2021}
G.~Xu, P.~Kordjamshidi, and J.~Chai, ``Zero-shot compositional concept
  learning,'' in \emph{Proceedings of the 1st Workshop on Meta Learning and Its
  Applications to Natural Language Processing (MetaNLP)}, 2021.

\bibitem{Xu_GIPCOL_WACV2024}
G.~Xu, J.~Chai, and P.~Kordjamshidi, ``{GIPCOL}: Graph-injected soft prompting
  for compositional zero-shot learning,'' in \emph{Proceedings of the IEEE/CVF
  Winter Conference on Applications of Computer Vision (WACV)}, 2024.

\bibitem{liu2024lookback}
Q.~Liu, A.~Singh, J.~Liu, C.~Mu, Z.~Yan, and J.~Pedersen, ``Long or short or
  both? an exploration on lookback time windows of behavioral features in
  product search ranking,'' \emph{arXiv preprint arXiv:2409.17456}, 2024.

\bibitem{exploitation_bias}
\BIBentryALTinterwordspacing
T.~Yang, C.~Han, C.~Luo, P.~Gupta, J.~M. Phillips, and Q.~Ai, ``Mitigating
  exploitation bias in learning to rank with an uncertainty-aware empirical
  bayes approach,'' in \emph{Proceedings of the ACM Web Conference 2024}, ser.
  WWW '24.\hskip 1em plus 0.5em minus 0.4em\relax New York, NY, USA:
  Association for Computing Machinery, 2024, p. 1486–1496. [Online].
  Available: \url{https://doi.org/10.1145/3589334.3645487}
\BIBentrySTDinterwordspacing

\bibitem{Xu_MetaReVision_EMNLP2023}
G.~Xu, P.~Kordjamshidi, and J.~Chai, ``{MetaReVision}: Meta-learning with
  retrieval for visually grounded compositional concept acquisition,'' in
  \emph{Findings of the Association for Computational Linguistics: EMNLP 2023},
  2023.

\bibitem{GBDT2001}
\BIBentryALTinterwordspacing
J.~H. Friedman, ``Greedy function approximation: A gradient boosting machine,''
  \emph{The Annals of Statistics}, vol.~29, no.~5, pp. 1189--1232, 2001.
  [Online]. Available: \url{https://www.jstor.org/stable/2699986}
\BIBentrySTDinterwordspacing

\bibitem{YDF}
\BIBentryALTinterwordspacing
M.~Guillame{-}Bert, S.~Bruch, R.~Stotz, and J.~Pfeifer, ``Yggdrasil decision
  forests: {A} fast and extensible decision forests library,'' in
  \emph{Proceedings of the 29th {ACM} {SIGKDD} Conference on Knowledge
  Discovery and Data Mining, {KDD} 2023, Long Beach, CA, USA, August 6-10,
  2023}, 2023, pp. 4068--4077. [Online]. Available:
  \url{https://doi.org/10.1145/3580305.3599933}
\BIBentrySTDinterwordspacing

\bibitem{lambdamart}
\BIBentryALTinterwordspacing
C.~J.~C. Burges, ``From {RankNet} to {LambdaRank} to {LambdaMART}: An
  overview,'' Microsoft Research, Tech. Rep., 2010. [Online]. Available:
  \url{https://www.microsoft.com/en-us/research/publication/from-ranknet-to-lambdarank-to-lambdamart-an-overview/}
\BIBentrySTDinterwordspacing

\bibitem{spark}
\BIBentryALTinterwordspacing
M.~Zaharia, R.~S. Xin, P.~Wendell, T.~Das, M.~Armbrust, A.~Dave, X.~Meng,
  J.~Rosen, S.~Venkataraman, M.~J. Franklin, A.~Ghodsi, J.~Gonzalez,
  S.~Shenker, and I.~Stoica, ``Apache spark: a unified engine for big data
  processing,'' \emph{Commun. ACM}, vol.~59, no.~11, p. 56–65, Oct. 2016.
  [Online]. Available: \url{https://doi.org/10.1145/2934664}
\BIBentrySTDinterwordspacing

\bibitem{gcp}
\BIBentryALTinterwordspacing
Google, ``Google cloud platform,'' accessed: 2026-05-12. [Online]. Available:
  \url{https://cloud.google.com/}
\BIBentrySTDinterwordspacing

\bibitem{tfserve}
\BIBentryALTinterwordspacing
C.~Olston, N.~Fiedel, K.~Gorovoy, J.~Harmsen, L.~Lao, F.~Li, V.~Rajashekhar,
  S.~Ramesh, and J.~Soyke, ``Tensorflow-serving: Flexible, high-performance ml
  serving,'' 2017. [Online]. Available: \url{https://arxiv.org/abs/1712.06139}
\BIBentrySTDinterwordspacing

\bibitem{redis}
\BIBentryALTinterwordspacing
Redis, ``Redis documentation,'' 2026. [Online]. Available:
  \url{https://redis.io/docs/}
\BIBentrySTDinterwordspacing

\bibitem{kafka}
J.~Kreps, N.~Narkhede, and J.~Rao, ``Kafka: a distributed messaging system for
  log processing,'' in \emph{Proceedings of the 6th International Workshop on
  Networking Meets Databases (NetDB '11)}, Athens, Greece, 2011.

\bibitem{Xu_PromptingVLM_CCL_arXiv2022}
G.~Xu, P.~Kordjamshidi, and J.~Chai, ``Prompting large pre-trained
  vision-language models for compositional concept learning,'' \emph{arXiv
  preprint arXiv:2211.05077}, 2022.

\bibitem{Xu_GroundedCCL_PhD2024}
G.~Xu, ``Grounded compositional concept learning,'' Ph.D. dissertation,
  Michigan State University, 2024.

\end{thebibliography}

\end{document}